\documentclass[
 reprint, amsmath,amssymb, aps,
]{revtex4-1}
\pdfoutput=1

\usepackage[colorlinks=true,linkcolor=black, citecolor=black,
urlcolor=black]{hyperref}

\usepackage{multirow,graphics}
    \newcommand{\beq}{\begin{equation}}
    \newcommand{\eeq}{\end{equation}}
    \newcommand\beqa{\begin{eqnarray}}
    \newcommand\eeqa{\end{eqnarray}}

\usepackage{amstext}
\usepackage{amssymb}
\usepackage{amsmath}
\usepackage{graphicx}

\usepackage[usenames,dvipsnames]{color}

\begin{document}

\newcommand{\IM}{{\rm Im}\,}
\newcommand{\card}{\#}
\newcommand{\la}[1]{\label{#1}}
\newcommand{\eq}[1]{(\ref{#1})}
\newcommand{\figref}[1]{Fig \ref{#1}}
\newcommand{\abs}[1]{\left|#1\right|}
\newcommand{\comD}[1]{{\color{red}#1\color{black}}}
\renewcommand{\d}{\partial}

\makeatletter
     \@ifundefined{usebibtex}{\newcommand{\ifbibtexelse}[2]{#2}} {\newcommand{\ifbibtexelse}[2]{#1}}
\makeatother

\preprint{Imperial/TP/13/SL/02}

\newcommand{\footnotea}[1]{\ifbibtexelse{\footnote{#1}}{
\newcommand{\textfootnotea}{#1}
\cite{thefootnotea}}}
\newcommand{\footnoteb}[1]{\ifbibtexelse{\footnote{#1}}{
\newcommand{\textfootnoteb}{#1}
\cite{thefootnoteb}}}
\newcommand{\footnotebis}{\ifbibtexelse{\footnotemark[\value{footnote}]}{
\cite{thefootnoteb}}}

\def\e{\epsilon}
     \def\bT{{\bf T}}
    \def\bQ{{\bf Q}}
    \def\wT{{\mathbb{T}}}
    \def\wQ{{\mathbb{Q}}}
    \def\ttQ{{\bar Q}}
    \def\tQ{{\tilde \bP}}
        \def\bP{{\bf P}}
    \def\CF{{\cal F}}
    \def\cC{\CF}
     \def\Tr{\text{Tr}}
     \def\l{\lambda}
\def\hbZ{{\widehat{ Z}}}
\def\bZ{{\resizebox{0.28cm}{0.33cm}{$\hspace{0.03cm}\check {\hspace{-0.03cm}\resizebox{0.14cm}{0.18cm}{$Z$}}$}}}
\newcommand{\rb}{\right)}
\newcommand{\lb}{\left(}
\newcommand{\gT}{T}\newcommand{\gQ}{Q}

\title{Exact Slope and Interpolating Functions  in ABJM Theory}

\author{ Nikolay Gromov$^{a,b}$, Grigory Sizov$^{a}$}

\affiliation{
\(^{a}\)Mathematics Department, King's College London, The Strand, London WC2R 2LS, UK
\\
\(^{b}\)
 St.Petersburg INP, Gatchina, 188300, St.Petersburg, Russia
               }

\begin{abstract}
Using the Quantum Spectral Curve approach we compute exactly an observable (called slope function) in
the planar ABJM theory in terms of an unknown interpolating function $h(\lambda)$ which plays the role of the coupling
in any integrability based calculation in this theory.
We verified our results with known weak coupling expansion in the gauge theory and
with the results of semi-classical string calculations. Quite surprisingly at strong coupling
the result is
given by an explicit rational function of $h(\lambda)$ to all orders.

By comparing the structure of our result with that of an exact localization based calculation for a similar observable in JHEP 1006 (2010) 011 we conjecture an exact expression
 for $h(\lambda)$.
\end{abstract}

 \maketitle

\section{Introduction}
The well known duality between ABJM theory in 3 dimensions and type IIA string theory in $AdS_4\times \mathbb{C}P_3$ is
an explicit example of $AdS/CFT$
 correspondence \cite{Aharony:2008ug}. The ABJM theory is ${\cal N}=6$ supersymmetric gauge theory with gauge group $U(N)\times U(N)$ consisting of two copies of super Chern-Simons theory at level $k$ and $4$ matter multiplets. A particularly interesting
 limit of ABJM theory is the planar limit, in which $N,k\rightarrow\infty$, whereas their ratio $\lambda=N/k$, called 't Hooft coupling, is fixed. In this limit ABJM  manifests signs of interability \cite{Klose:2010ki}, as was first noticed in \cite{Minahan:2008hf}.
 This feature allows for completely non-perturbative calculations in this fully interacting non-abelian gauge theory
 as we will explore further and exemplify in this paper.

 ABJM theory has many features similar to ${\cal N}=4$ SYM, but there are important differences.
 Both have rather similar integrability structures, but whereas in
 ${\cal N}=4$ SYM integrability gives the result in terms of the `t Hooft coupling $\lambda$,
in ABJM theory the predictive power of integrability is limited due to one unknown
function $h(\lambda)$ entering into all integrability based calculations. This function is only known in the weak and strong coupling expansions.
In this paper we also conjecture an exact form of this function based on indirectly comparing our all-loop results with localization calculations.

\section{Quantum Spectral Curve}
In this section we describe the Quantum Spectral Curve (QSC) also known as $\bP\mu$-system for ABJM model of \cite{CFGT}.
The structure found in \cite{CFGT} has an unexpected and intriguing relation
to that of QSC of ${\cal N}=4$ SYM proposed by \cite{Gromov:2013pga,LargePaper}.
Here we briefly describe the part of the construction essential for our applications.

The main objects are $5$ functions ${\bf P}_A,\;A=1,\dots,5$
and $4$ function $\nu_a,\;a=1,\dots,4$ of the spectral parameter $u$.
${\bf P}_A$ are restricted by a quadratic constraint $\bP_5=\sqrt{1-\bP_1\bP_4+\bP_2\bP_3}$.
Depending on the choice of the branch cuts $\nu_a$ could be made $i$-periodic.
However, for our calculation it will be more convenient to choose the branch of $\nu_a(u)$
with infinitely many cuts going from $-2h+i n$ to $2h+i n$
for any integer $n$. In this case $\nu_a$ are quasi-periodic and satisfy
$\nu_a(u+i)=-{\bf P}_{ab}(u)\nu^b(u)$, where $\nu^1=-\nu_4,\;,\nu^2=\nu_3,\;\nu^3=-\nu_2,\;\nu^4=\nu_1$ and ${\bf P}_{ab}$ is a $4\times 4$
matrix built out of ${\bf P}_A$:
\beq\la{Pab}
{\bf P}_{ab}=
\left( \begin{array}{cccc} 0 & - {\bf P}_1 & -{\bf P}_2 & -{\bf P}_5 \\ {\bf P}_1 & 0 & -{\bf P}_5  & -{\bf P}_3 \\ {\bf P}_2 & {\bf P}_5 & 0 & -{\bf P}_4 \\ {\bf P}_5 & {\bf P}_3 &  {\bf P}_4 & 0 \end{array} \right)_{ab}\;.
\eeq
Analytical continuation of $\nu_a$ under the cut $[-2h,2h]$, denoted as $\tilde \nu_a$,
is related to $\nu_a$ itself simply by $\tilde \nu_a(u)=\nu_a(u+i)$.
Finally, functions ${\bf P}_A$ have only one cut $[-2h,2h]$
and their analytical continuation $\tilde\bP_A$ under this cut is given by
\beq
\tilde \bP_{ab}=\bP_{ab}+\nu_a\tilde\nu_b-\nu_b\tilde\nu_a\;.
\label{Pmonodromy}
\eeq

We note that this construction is very similar to that of ${\cal N}=4$ SYM:
indeed, replacing ${\bP}_{ab}$ by $\mu_{ab}^{{\cal N}=4}$ and $\nu_a$ by $\bP_a^{{\cal N}=4}$,
algebraically, we get exactly the same equations. However, their analytical properties are interchanged (see \cite{CFGT} for more details).

Finally, we have to specify how the quantum numbers of a state enter into this construction.
In this letter we focus on $sl_2$ subsector, which includes single trace operators of the type
${\rm tr}[D_+^S (Y^1 Y_4^\dagger)^L]$ \footnote{Strictly speaking these operators could also mix with fermions, for a detailed description see \cite{Klose:2010ki}}, thus there are $3$ quantum numbers to specify -- $L,S$ and the scaling dimension
$\Delta=L+S+\gamma$, where $\gamma$ denotes its anomalous part.
They all enter the $\bP\mu$-system through the large $u$ asymptotics
\beq
\bP_A\sim (A_1 u^{-L},A_2 u^{-L-1},A_3 u^{+L+1},A_4 u^{+L},A_5 u^0)
\label{Pasympt}
\eeq
and $\Delta$ and $S$ are encoded into the coefficients as
\beqa\la{As}
A_1A_4&=&-\frac{2 \left(({\bf L}-S)^2-{\bf \Delta}^2\right) \left(({\bf L}+S-1)^2-{\bf \Delta }^2\right)}{(1-2 {\bf L})^2 {\bf L}},\\
\nonumber A_2A_3&=&-\frac{2 \left(({\bf L}+S)^2-{\bf \Delta}^2\right)
   \left(({\bf L}-S+1)^2-{\bf \Delta}^2\right)}{ (1+2 {\bf L})^2{\bf L}},
\eeqa
where we introduced ${\bf \Delta}=\Delta+\tfrac12$ and ${\bf L}=L+\tfrac12$.

\begin{figure}
  \includegraphics[scale=0.2]{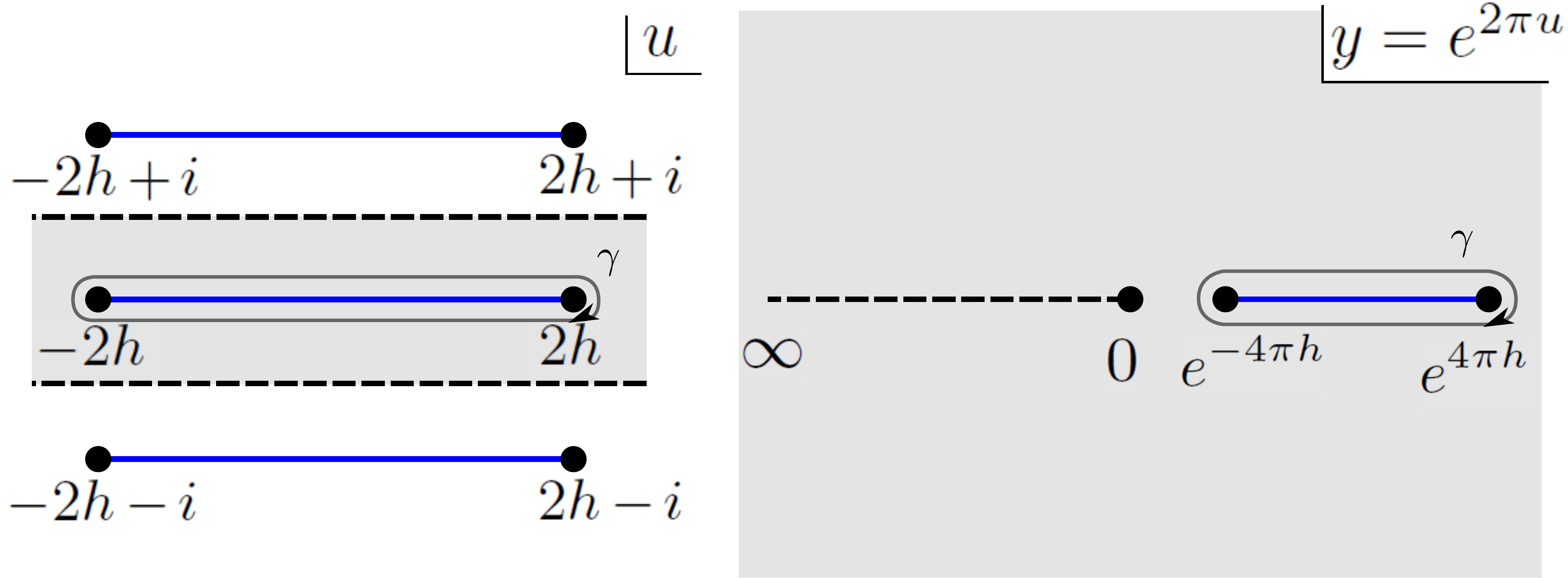}
  \caption{Antiperiodic functions $\rho_{2,3}(u)$ have infinitely
  many short cuts. In the variable $y=e^{2\pi u}$ only two cuts remain.
  }
  \label{fig:cuts}
\end{figure}
\section{Slope Function}
In this letter we compute the slope function exactly as a function of the effective coupling $h(\lambda)$.
This observable is close to the BPS point in the parameter space.
Similar observables were studied in ${\cal N}=4$ SYM and an enormous simplification of the $\bP\mu$ system equations was observed there, which allowed for an exact explicit solutions for any value of
`t~Hooft coupling.
For us the BPS operator is ${\rm tr}[(Y^1 Y_4^\dagger)^L]$ so in order to be close to the protected point we have to take the number of derivatives $S$
 small. The expansion coefficients in small $S$ are expected to be exactly computable, and we will find the first such coefficient, called slope function.
 In ${\cal N}=4$ SYM it was first computed by Basso \cite{Basso:2011rs}. There the situation was {\it a priori} simpler as in ${\cal N}=4$
 the slope function is not
 affected by wrapping effects which means that the result can be calculated solely from a simple algebraic set of equations called asymptotic Bethe ansatz. At the same time
 in the QSC formalism the wrapping corrections are incorporated automatically and both theories can be treated very similarly.
 As we will show in many ways the calculation in ABJM based on QSC is similar to that of \cite{Gromov:2014bva} in ${\cal N}=4$ SYM.

To see that there is a simplification in the limit $S\rightarrow 0$, we notice that in this limit $\Delta\simeq L$ and thus $A_1A_4\sim A_2A_3\sim S$.
Like in ${\cal N}=4$ case this indicates that $\bP_a\sim\sqrt S,\;a=1,\dots,4$ and due to the constraint we must have
 $\bP_5\simeq 1$. Based on this, we found that the consistent scaling of $\nu_a$ is $\nu_1,\nu_4\sim1$, $\nu_2,\nu_3\sim\sqrt S$.
 Then the leading order of the equations for the monodromy of $\nu_a$ becomes
\beq
\left(\begin{array}{c}
  \tilde \nu_1 \\
  \tilde \nu_2 \\
  \tilde \nu_3 \\
  \tilde \nu_4
\end{array}\right)=\left(\begin{array}{cccc}
   1 & 0 & 0  & 0 \\
   \bP_3 & -1 & 0 & \bP_1   \\
   \bP_4 &      0 & -1 & \bP_2  \\
   0     & 0 & 0 & 1
\end{array}\right)\left(\begin{array}{c}
\nu_1 \\
 \nu_2 \\
 \nu_3 \\
 \nu_4
\end{array}\right)\;,
\label{eqnu}
\eeq
from where we see that to the leading order $\nu_1$ and $\nu_4$ do not have cuts whereas $\nu_2$ and $\nu_3$ are nontrivial.
There is still certain freedom in the construction which allows, for example,
to shift $\bP_4\rightarrow \bP_4+\alpha \bP_2$ with arbitrary constant $\alpha$. We use this freedom to set $\nu_1=1$
and $\nu_4=0$ \cite{CFGT}. The equations for the monodromy of $\bP_a$ take the form
\beqa
&&\bP_1-\tilde \bP_1=\tilde\nu_2-\nu_2,\  \tilde \bP_3=\bP_3,\nonumber \\
&& \bP_2-\tilde \bP_2=\tilde\nu_3-\nu_3,\ \tilde \bP_4=\bP_4.
\label{eqP}
\eeqa
From \eqref{eqnu}, \eqref{eqP} one can see that the equations for $\bP_2,\bP_4,\nu_3$ decouple from $\bP_1,\bP_3,\nu_2$. The two groups of equations differ only by asymptotics of $\bP_a$, so here we only give details on the solution of the first one, i.e.
$
\tilde \nu_3+\nu_3=\bP_4,\; \bP_2-\tilde \bP_2=\tilde\nu_3-\nu_3,
$
which together with periodicity gives $\nu_3(u+i)+\nu_3(u)=\bP_4(u)$. There is another important difference between $\nu_2$ and $\nu_3$ -- we have to assume that
asymptotics of $\nu_3$ grows as $e^{\pi u}$ and $\nu_2$ decays at infinity.
This is a peculiarity of analytical continuation in $S$ to non-integer values, described
for ${\cal N}=4$ SYM in
 \cite{Gromov:2014bva} in detail.

Taking into account that $\bP_4$ does not have a cut according to \eq{eqP} and also its asymptotic behavior \eqref{Pasympt}, we conclude that it is a polynomial in $u$ of degree $L$.
Introducing notations
$
\nu_3(u+\tfrac i2)=\rho_3(u+\tfrac i2)+Q_3(u)\;\;,\;\;
$
where $Q_3(u)$ is a polynomial such that $Q_3(u+\tfrac i2)+Q_3(u-\tfrac i2)=\bP_4(u)$,
we get
\beq\la{tosolve}
\rho_3(u+i)=-\rho_3(u)\;\;,\;\;\tilde\rho_3+\rho_3=
Q_3^+-Q_3^-\equiv q_3\;,
\eeq
i.e. $\rho_3$ is antiperiodic. It is convenient to make a change of variables $y=e^{2\pi u}$, which maps infinitely many cuts of $\nu_i$ or $\rho_i$ into one cut and introduces a quadratic cut from $0$ to $-\infty$, see Fig. \ref{fig:cuts}.

In order to resolve equations of the form $\tilde g+g=f$ like in \eq{tosolve} we define the following Hilbert transformation $H$ as
\beq\la{hilbert}
H[f](z)=\frac{1}{2}\oint_\gamma\frac{dy}{2\pi i} \frac{\sqrt{z-e^{4\pi h}}\sqrt{z-e^{-4\pi h}}}{\sqrt{y-e^{4\pi h}}\sqrt{y-e^{-4\pi h}}}\frac{f(y)}{y-z},
\eeq
which gives a solution of this equation with non-growing asymptotics at infinity with one cut $[e^{-4\pi h},e^{+4\pi h}]$.
Note, however, that $\rho_3(z)$ has another cut $(-\infty,0)$ on which it simply changes its sign.
We can overcome this problem by dividing it by $\sqrt z$. After that we can use \eq{hilbert} to get
\beq
\rho_3(z)=\frac{C}{\sqrt{z}}+{\sqrt{z}}H\left[\frac{1}{\sqrt{y}}
q_3\left(\frac{\log y}{2\pi}\right) -\frac{2C}{ y}\right]\;.
\eeq
The term proportional to $C$ is added here, because it is not prohibited by the asymptotics:
 $\nu_3$ can grow as $e^{\pi u}$ at infinity.
 The constant $C$ is fixed at the end from the condition that $\rho_3(u)$ should be even.

Next, knowing $\rho_3$ and thus $\nu_3$ in terms of the yet to be fixed polynomial $\bP_4$,
we can find $\bP_2$ as a solution of corresponding equation in \eqref{eqP}. As $\bP_2$ is a function with one cut we simply
use the Cauchy kernel
\beq
\bP_2(v)=-\oint \frac{dz}{2\pi i z}\frac{\rho_3(z)}{\log z-2\pi v}.
\eeq

Thus we found all the objects in terms of a few coefficients of the polynomial $\bP_4$.
To extract $\gamma_L(h)$ we have to find these coefficients. Consider first, for simplicity, $L=1$.
 In this case  $\bP_4=A_4 u$ , so $q_3=i\tfrac{A_4}{2}$ and $\bP_2=A_4\oint \frac{dz}{4\pi \sqrt{z}}\frac{H[y^{-1/2}]}{2\pi v-\log z}$.
 Thus considering the leading asymptotics of $\bP_2$ we can obtain $A_2/A_4=
\oint \frac{dz \log{z}}{2(2\pi)^3\sqrt{z}}H[y^{-\frac{1}{2}}]$  and similarly $\bP_1$ gives $A_1/A_3$.
On the other hand, expanding equations \eqref{As} to the first order in $S$ yields $\gamma_1=\frac{-2S}{1+\frac{L}{L+1}\frac{A_1 A_4}{A_3 A2}}$ and
substituting the ratios of the coefficients we get
\beq
\gamma_1(h)=-2S\frac{{\d_\alpha I_{-\frac12,-\frac12}}}{{\d_\alpha I_{-\frac12,-\frac12}+{\d_\beta I_{-\frac12,-\frac12}}}},
\eeq
where
\beq\la{Iab}
I_{\alpha,\beta}=\oint_\gamma {dy}{} \oint_\gamma  {dz} \frac{\sqrt{y-e^{4\pi h}}\sqrt{y-e^{-4\pi h}}}{\sqrt{z-e^{4\pi h}}\sqrt{z-e^{-4\pi h}}}
\frac{ y^\alpha z^\beta}{z-y}.
\eeq
Both integrals go around the cut $[e^{-4\pi h},e^{4 \pi h}]$. Another convenient representation of
$I_{\alpha,\beta} e^{4 \pi  h (\alpha +\beta +1)}$ is
\beq\la{convenient}\!\pi ^2{\beta}\!\!\!\!\int\limits_0^{e^{8\pi h}-1}\!\!\!\!\!\!
   _2F_1\left(\tfrac{3}{2},-\alpha ;2;-S\right)\!{}_2F_1\left(\tfrac{3}{2},1-\beta ;2;-S\right)\!S dS.
\eeq

 For odd $L>1$ there are $\frac{L+1}{2}$ constants in $q_3$. To fix them we use $\frac{L-1}{2}$ conditions of the form
 $\oint \frac{du}{2\pi i }u^{k}\gamma_3(u)=0$ for $k=1,3\dots L-2$,
  which ensure
 that asymptotics of $\bP_2$ at infinity is ${\cal O} \left(u^{-L-1}\right)$. These conditions take a form of a system of linear equations for constants entering $q_3$ with coefficients of the form
$s_{n,k}=\partial^k_\alpha\partial^n_\beta I_{-1/2,-1/2}$. The solution for this system takes form of a ratio of determinants made of $s_{n,k}$.
Similar strategy also applies to $\bP_1, \bP_3$ and $\nu_2$.

Using again formula for $\gamma_L$ in terms of $A_1/A_3$, $A_2/A_4$ we get that the answer for any $L$ \footnote{the procedure for odd $L$, which we do not describe here, is analogous.} is
\beq\la{result}
\gamma_{L}=-\frac{2S}{1+  r_L/ r_{L-2}}\;,
\eeq
where
\beq
 r_L=\frac{\det s_{L-2i-1,L-2j}}{\det s_{L-2i,L-2j-1}},\ i,j=0\dots \lfloor\tfrac{L}{2}\rfloor,\ L\ge0
\eeq
$s_{k,n}=\partial^k_\alpha\partial^n_\beta I_{-1/2,-1/2}$ for $k,n\ge0$,  for negative indexes we define
$s_{k,-1}= \partial_\alpha^k I_{-1/2,-1}$ and
$s_{-1,n}=\frac{1}{2}\partial_\beta^n \left.\left(e^{4\pi\beta h}{}_2F_1\left(\frac{1}{2},\!-\beta;1,1\!-\!e^{8\pi h}\right)\right)\right|_{\beta=-1/2}$.

Equation \eq{result} is our result for the slope function which we now test at weak and strong coupling.

\paragraph{Weak coupling.}
At weak coupling it is convenient to use \eq{convenient}.
Up to the order $h^{2L}$ we can compare our result against the
slope function $\gamma_L^{ABA}=S\frac{2\pi h}{L}\frac{I_{J+1}(2\pi h)}{I_J(2\pi h)}$ of ${\cal N}=4$ SYM \cite{Basso:2011rs}
which does not take into account the wrapping effects.
These effects appear at the order ${\cal O}\left(h^{2L+2}\right)$ and the leading
deviation can be compared with the Luscher correction
which we found as a generalization of \cite{Gromov:2009tv,Beccaria:2009ny}:
\beq
\gamma^{wrap}_L=-Sh^{2L+2}\frac{4\pi^{3/2}(4^L-2)\zeta_{2L}\Gamma\left(L+\frac{1}{2}\right)}{L\Gamma(L+2)}.
\eeq
We found a perfect agreement with our exact formula for $L=1,\dots,5$ and $7$.

\paragraph{Strong coupling.}
At strong coupling we notice an interesting phenomenon -- our result can be written
explicitly as a rational function of $h$ with exponential precision.
For example
$\gamma_{L=1}=\frac{4{\bf g}^3-12 {\bf g}^2+12{\bf g}-3\zeta_3}{6{\bf g}^2-6{\bf g}}S+{\cal O}(e^{-4{\bf g}})
$
where ${\bf g}=2\pi h+\log 2$.
To get this expression
we have evaluated the integral \eq{Iab} with exponential precision
\beqa
&&I_{\alpha\beta}\simeq-\frac{2 \pi  \Gamma \left(\alpha +\frac{1}{2}\right) \Gamma \left(-\beta
   -\frac{1}{2}\right) e^{4 \pi  h (\alpha -\beta )}}{\Gamma (\alpha +2) \Gamma
   (-\beta )}\\
\nonumber&&-\frac{4 \pi  \left(\beta +\frac{1}{2}\right) \Gamma \left(-\alpha
   -\frac{1}{2}\right) \Gamma \left(-\beta -\frac{1}{2}\right) e^{4 \pi  h
   (-\alpha -\beta -1)}}{(\alpha +\beta +1) \Gamma (-\alpha ) \Gamma (-\beta
   )}\\
\nonumber&&+\frac{4 \pi  (\alpha+\tfrac12)\Gamma \left(+\alpha +\frac{1}{2}\right) \Gamma
   \left(+\beta +\frac{1}{2}\right) e^{4 \pi  h (+\alpha +\beta +1)}}{(\alpha
   +\beta +1) \Gamma (\alpha +2) \Gamma (\beta )}\;.
\eeqa
We found that for any $L$ the result is some rational function of ${\bf g}$ of a growing with $L$ complexity.
However, the large ${\bf g}$ expansion coefficients can be found explicitly for any $L$ to be
\beq\la{slopestrong}
\frac{\gamma_{L}}{S}=
\frac{\mathbf{g}-L-1}{L+\frac12}
+\left(\frac{1}{2\mathbf{g}}
   -\frac{3 \zeta_3-4}{8\mathbf{g}^2}\right) \frac{L^2+L}{ L+\frac12}+{\cal O}({\bf g}^{-3}).
\eeq
To test our result we take the quasi-classical limit $L\sim {\bf g}\gg1$.
Introducing
$
{{\cal J}}=\frac{L+1/2}{{\bf g}}\sim 1
$
and ${\cal S}=\frac{S}{\bf g}$ and expanding at large ${\bf g}$ in \eq{slopestrong}
we find
\beqa
&&\frac{\Delta-L}{S}\simeq{\left(
\frac{1}{\mathcal{J}}+\frac{\mathcal{J}}{2}+\dots\right) }+
\nonumber\frac{1}{\bf g}\left(\frac{-1}{2 \mathcal{J}}+
    \mathcal{J}\frac{4-3 \zeta _3}{8} +\dots\right)
\eeqa
which reproduces the corresponding terms in the tree level and one-loop quasi-classical
folded string quantization \cite{Beccaria:2012vb}. Note that with our
definition of ${\cal J}$ all $\log2$ terms and all even powers of ${\cal J}$
disappear from the one-loop terms of \cite{Beccaria:2012vb}.
From that we can see that
${\bf L}\equiv L+1/2$, which appears in denominator of \eq{slopestrong}, and ${\bf \Delta}\equiv \Delta+1/2$ are natural
combinations as is already clear at the level of \eq{As} which only depends on ${\bf \Delta}^2$
and where under the change of sign of ${\bf L}$ the two lines in \eq{As} simply interchange.
This hints the following ansatz for double expansion at large ${\bf g}$ and small $S$, similar to the result of \cite{Basso:2011rs} for ${\cal N}=4$ SYM
\beq\la{ansatz}
{\bf \Delta}^2-{\bf L}^2=\sum_{n,k=1} A_{n,k}({\bf L}^2)S^n {\bf g}^{-n-k+3}
\eeq
where the coefficients $A_{n,k}$ are polynomials of degree $\lfloor \tfrac k2\rfloor$ in ${\bf L}^2$.
By comparison with \eq{slopestrong} and
with quasi-classics \cite{Beccaria:2012qd}
we find $A_{1,1}=2$, $A_{1,2}=-1$
, $A_{1,3}={\bf L}^2-\tfrac14$,
$A_{1,4}=({\bf L}^2-\tfrac14)(1-\tfrac{3\zeta_3}{4})$,
$A_{2,1}=\tfrac{3}{2}$, $A_{2,2}=\tfrac{5}{8}-\tfrac{9\zeta_3}{4}$. Next we can re-expand \eq{ansatz} sending ${\bf g}\to\infty$ like in \cite{Basso:2011rs}.
For example at $L=1,S=2$ we get
\eq{ansatz}
\beq
\Delta_{L=1,S=2}=2\sqrt{\bf g}-\frac{1}{2}+\frac{25}{16\sqrt{\bf g}}
+\left(\frac{271}{1024}-\frac{9\zeta_3}{4}\right){\bf g}^{-3/2}+\dots\;.
\eeq
which gives a prediction for a strong coupling expansion of the anomalous
dimension of a short operator. As we see this result can be trivially generalized to any $S$ and $L$, but the expression we found are rather bulky.
We also note that the third term disagrees with \cite{Beccaria:2012qd}, which is most likely due to the different ansatz used in \cite{Beccaria:2012qd}.
As our ansatz is based on an extra insight about the structure of the spectrum coming from QSC and the symmetries of \eq{As}
our result is likely to be the correct one. It would be interesting to use
the methods of \cite{Frolov:2013lva} to check this result.
That is important to note that it is not expected that this result holds for odd $S$, as operators with
odd $S$ belong to a different trajectory, 
as can be seen already at weak coupling \cite{Zwiebel:2009vb,Beccaria:2009wb}. In particular the analytical continuation of $\gamma$ from odd $S$ does not go through the BPS point
and does not vanish at $S=0$ and thus should be treated differently \footnote{We thank B.Basso for pointing this subtlety out to us.}.

\section{Comparison with Localization}
Here we compare the structure of our result for the slope function with the result of
\cite{Marino:2009jd,Drukker:2011zy,Drukker:2011zyprime} obtained using localization \cite{Pestun:2007rz,Kapustin:2009kz}. The quantity calculated in \cite{Marino:2009jd} is the expectation value of $1/6$ BPS Wilson loop, which in ${\cal N}=4$ is known to similar to the slope function. Although in ABJM these quantities are not related that closely, we still expect similarity in structure, which allow us to make a conjecture about $h(\lambda)$.
The result of \cite{Marino:2009jd} can be written in a parametric form in terms of $\kappa$
as an integral over the matrix-model eigenvalue $\log Z$
\beq\la{W16}
\langle
W^{1/6}_{m=1}
\rangle
=
\int_{\frac1{A^+}}^{A^+} \frac{dZ}{2\pi^2 i\lambda}
 \arctan\sqrt\frac{2+i\kappa-Z-\frac{1}{Z}}{2-i\kappa+
Z+\frac{1}{Z}}
\eeq
we see that the argument of $\arctan$ has $4$ branch-points.
The integration goes between the branch-points from the
numerator are $A^+$ and $1/A^+$ and those from the denominator, which we
denote $A^-$ and $1/A^-$ where
\beq
A^\pm = \pm\tfrac{1}{2}\left(2\pm i\kappa+\sqrt{\kappa(\pm 4i-\kappa)}\right)
\eeq
and the parameter $\kappa$ is related to the `t Hooft coupling by \cite{Marino:2009jd}
$\lambda=\frac{\kappa}{8\pi}{}_3F_2\left(\frac{1}{2},\frac{1}{2},\frac{1}{2};1,\frac{3}{2};-\frac{\kappa^2}{16}\right)
$.
The main observation is that the integral \eq{W16} is similar to the main ingredient of our result
\eq{Iab}. To make the similarity more clear, one can make a change of variable with a suitable
Mobius transformation which will map the branch points
$A^-\to\infty, 1/A^-\to 0$ and $A^+\to G, 1/A^+\to 1/G$ like on Fig.\ref{fig:cuts}.
There is a unique Mobius transformation with this property. Furthermore, it fixes uniquely the value of $G$ in terms of $\kappa$ as
$G=\left(\tfrac14 (\sqrt{\kappa^2 + 16} + \kappa)\right)^2
$, which is easy to find
from the cross-ratio of the branch points before and
after the transformation.
Thus to relate \eq{W16} with \eq{Iab} we set $G=e^{4\pi h}$
which leads to our conjecture
\beq
\lambda=\frac{\sinh(2\pi h)}{2\pi}{}_3F_2\left(\frac{1}{2},\frac{1}{2},\frac{1}{2};1,\frac{3}{2};
-\sinh^2(2\pi h)\right)\;.
\label{conj}
\eeq
Expansion at weak/strong coupling gives
\beqa
h(\lambda)&=&\lambda-\frac{\pi ^2
   \lambda ^3}{3}+\frac{5 \pi ^4
   \lambda ^5}{12}-\frac{893 \pi ^6 \lambda
   ^7}{1260}+{\cal O}(\lambda^9),\nonumber\\
h(\lambda)&=&
\sqrt{\frac{1}{2}\left(\lambda-\frac{1}{24}\right)
   }-\frac{\log 2}{2 \pi }+{\cal O}\left(
   e^{-
    \pi  \sqrt{8\lambda
   }}\right),
\eeqa
which reproduces all known coefficients at weak
and at strong coupling i.e. in total $4$ nontrivial coefficients \cite{Minahan:2009aq,Leoni:2010tb,McLoughlin:2008he,Abbott:2010yb,LopezArcos:2012gb}. Curiously, the shift by $-\frac{1}{24}$ at strong coupling coincides with the anomalous radius shift of AdS found in \cite{Bergman:2009zh}, as also noticed in \cite{Drukker:2011zyprime}.

Of course such identification at the level of the integrands is not completely rigorous and in order to derive $h(\lambda)$ one
should apply the method of the QSC to the Bremsstrahlung function like in \cite{Drukker:2012de,Correa:2012hh,Gromov:2012eu,Gromov:2013pga}
and compare it to the result from localization \cite{Drukker:2011zy,Lewkowycz:2013laa,Agon:2014rda} for {\it the same} quantity (for recent results on weak and strong coupling expansions of Bremsstrahlung function see \cite{Forini:2012bb,Bianchi:2014laa}).
But at the same time there are rather clear indications that we snatched the correct result.

\section{Summary}
In this letter we have applied the Quantum Spectral Curve \cite{Gromov:2013pga} method developed in \cite{CFGT} for ABJM to calculation of the exact slope function in this theory. Our result \eqref{result} has been checked to agree with the existing predictions at weak and strong coupling.
Our computation provides a highly non-trivial test of the QSC of \cite{CFGT} in ABJM.
 Also we proposed an ansatz for the anomalous dimension of short operators which allowed us to get the first $4$ nontrivial expansion coefficients.
We note that, similar to what was found in \cite{Gromov:2012eu}, the slope function is expressed through the ratio of determinants,
so it can be obtained as an expectation value in some matrix model (different from those arising in localization).
It would be interesting to investigate the fundamental role of these matrix models arising in the near BPS limit.

Comparing the structure of our result \eqref{result}
with that of a localization calculation for a different, but closely related observable -- $1/6$ BPS Wilson loop,
 we were able to conjecture the exact expression \eqref{conj} for $h(\lambda)$.
 On this way we also identified the relation between the eigenvalues of the localization matrix model and
 the spectral parameter. We can speculate that such relation indicates existence of a more general unifying structure which described
 non-BPS and non-planar physics combining nice features of localization and integrability methods. In this hypothetical description the usual for integrability
 Zhukovsky cuts would get discretized by the eigenvalues at finite $N$'s. It would be interesting to see whether such interpretation is indeed possible.

  \begin{acknowledgments}
\label{sec:acknowledgments}
We thank
B.~Basso, A.~Cavaglia, N.~Drukker, D.~Fioravanti, F.~Levkovich-Maslyuk, R.~Roiban, R.~Tateo, A.~Tseytlin and
S.~Valatka for discussions and useful comments on the draft.
The research of N.G. and G.S. leading to these results has received funding from the
People Programme (Marie Curie Actions) of the European Union's Seventh
Framework Programme FP7/2007-2013/ under REA Grant Agreement No 317089.
\end{acknowledgments}

\end{document}